\documentstyle[12pt]{article}

\newcommand{\ue}%
{\mbox{c\hspace{-0.4em}\rule[ 0.5ex]{0.3em}{0.04ex}\hspace{0.1em}}}
\newcommand{\Ue}%
{\mbox{C\hspace{-0.6em}\rule[ 0.75ex]{0.4em}{0.04ex}\hspace{0.2em}}}
\newlength{\signlength}%
\newcommand{\gs}%
{\settowidth{\signlength}{$<$}%
\raisebox{-0.35\signlength}%
{\makebox[1.7\signlength][c]{\makebox[0.pt]{$\sim$}%
\raisebox{0.6\signlength}{\makebox[0.pt]{$>$}}}}}
\newcommand{\ls}%
{\settowidth{\signlength}{$<$}%
\raisebox{-0.35\signlength}%
{\makebox[1.7\signlength][c]{\makebox[0.pt]{$\sim$}%
\raisebox{0.6\signlength}{\makebox[0.pt]{$<$}}}}}
\renewcommand{\ge}%
{\settowidth{\signlength}{$>$}%
\setlength{\unitlength}{0.1\signlength}%
\mbox{\begin{picture}(17,7.5)%
\linethickness{0.045\signlength}%
\put(0.0,1){\makebox(17,7.5){$>$}}%
\multiput(4.5,-1)(0.25,0.125){32}{\line(1,0){0.25}}%
\end{picture}}}
\renewcommand{\le}%
{\settowidth{\signlength}{$>$}%
\setlength{\unitlength}{0.1\signlength}%
\mbox{\begin{picture}(17,7.5)%
\linethickness{0.045\signlength}%
\put(0.0,1){\makebox(17,7.5){$<$}}%
\multiput(12.,-1)(-0.25,0.125){32}{\line(-1,0){0.25}}%
\end{picture}}}
\title{Uemura plot as a certificate of
two-dimensional character of superconducting transition for
quisi-two-dimensional HTS}
\author{G.Sergeeva}
\date{}
\begin{document}
\maketitle
\thispagestyle{plain}

\begin{center}

National Science Center "Kharkov Institute of Physics and
Technology",
Academicheskaya st.1, 61108, Kharkov, Ukraine

E-mail: gsergeeva@kipt.kharkov.ua

\end{center}

\begin{abstract}

For quisi-two-dimensional HTS in
superconducting state the dimensional
crossover, $3D \rightarrow 2D$ is
studied. With using general properties
of superconducting state for 2D systems
the universal temperature dependence of
the relation of the penetration lengths
of magnetic field along axis
$\widehat {c}$,
$\lambda^{2}(0)/\lambda^{2}(T/T_c)$, is
found out. This dependence evidences
about two-dimensional character of
superconducting transition for
quisi-two-dimensional HTS and leads to
the Uemura plot for quisi-two-dimensional HTS.
\end{abstract}

{\bf 1. Introduction}.
Quisi-two-dimensional (quisi 2D) or highly anisotropic
underdoped HTS are exhibiting such out of the ordinary
properties as "semiconducting-like" c-axis resistivity,
$\rho_c(T)$, near $T_c$, big interval of
two-dimensional superconducting fluctuations,
$\Delta^{N}_{2D} \sim T_c$, and pseudo-gap states
at $T\le T^{*}$, where $T^{*}$
is the temperature of charge ordering. These
properties lead to the discussion of the association of
superconducting transition with
Berezinskii-Kosterlitz-Thouless (BKT) transition
in $Cu O_2$ planes at $ T_{BKT}< T_{c0}$, where
$ T_{c0}$ is the temperature of two-dimensional
superconducting transition in the mean-field theory
(see, for example, review [1] and the references there).
Despite the observations of two-dimensional character of
superconducting fluctuations and the evidences for a BKT
transition have been reported in most of quisi
2D HTS [2,3], up to now the question whether
a BKT transition is observable in bulk cuprates
with taken into account interlayer coupling
in these materials [4,5] is under the discussion.

Interlayer coupling can lead to the dimensional crossovers
both as at $T \ge T_c$ ($2D \rightarrow 3D$), so in
superconducting state near $T_c$ ($3D \rightarrow 2D$)
[4-6]. Theoretical model of such superconducting
transition is well known [7-9]: at enough small
probability $t_c$ of charge tunnelling between
$Cu O_2$ planes transition has two-dimensional
character with finite region $\Delta_{3D}$ of
three-dimensional superconducting fluctuations.
Anisotropy of exchange interactions in $Cu O_2$
planes and along axis $\widehat {c}$ in spin-fluctuational
pairing model assumes big values of $\Delta^{N}_{2D}$ and
$T_{c0}$ and the distinction of temperatures
$T_c<<T_{c0}$. For quisi 2D HTS in ref.
[10,11] it was shown that at $T>T_c$ two-dimensional
character of superconducting fluctuations leads to
the temperature dependence of the probability of
charge tunnelling $t_c(T)$, and to the exasperation
of semiconducting-like of $\rho_c(T)$ near $T_c$, so
that $2D \rightarrow 3D$ crossover occurs at
$T^{0}_c>T_c>T_{BKT}$ before BKT transition. This
point out to the two-dimensional character of
superconducting transition, which develops
under Kats scenario [7] at enough small value of
 probability $t_c(T^{0}_c)$:
\begin{equation}\label{1}
T^{0}_c/ \varepsilon _F < t_c(T^{0}_c),
\end{equation}
where $\varepsilon _F$ is Fermi energy, values
$T^{0}_c$ and the temperature of three-dimensional
transition in the mean field theory are the same
order values. The interval $\Delta^{N}_{3D}$ of
three-dimensional fluctuations in normal state
can be found out from the measurements of resistivity
$\rho_c(T)$
\begin{equation}\label{2}
 \Delta^{N}_{3D} \simeq T^{0}_c- T_c << \Delta^{N}_{2D}
\end{equation}

It is known that at $T<T_c$ back crossover
$3D \rightarrow 2D$ occurs at $T_{cr}$ which value
depends on correlation length $ \xi _c (T)$ along
axis $\widehat {c}$ [6, 12]. This paper
is devoted to the studying of the
dimensionality of the superconducting state of quisi
2D HTS and to the determination of the values $T_{cr}$.
With using general properties of superconducting state
for 2D systems the universal temperature dependence of
the relation of the penetration lengths of magnetic
field along axis $\widehat {c}$,
$\lambda^{2}(0)/\lambda^{2}(T/T_c)$, is found out.

{\bf 2. Universal dependence of $T_c(\lambda^{-2}(0))$}.
The penetration lengths of magnetic field along axis
$\widehat{c}$, $\lambda(T)$ determines by London formula
$\lambda(T)\simeq n^{-1/2}_{s,3}$, where $n_{s,3}$ is the
three-dimensional superfluid density. The density
$n_{s,3}$ evidently interconnects with two-dimensional
superfluid density $n_{s,3}= n_s (T)\nu/l $, where
$\nu$ is number of layers, $l$ is the lattice
constant. V.Pokrovskii shown that for quisi 2D HTS
the penetration length $\lambda(T)$ and
two-dimensional superfluid density $n_s (T)$ is
connected by the expression
\begin{equation}\label{3}
\lambda^{2}(0)/ \lambda^{2}(T)=n_{s,3}(T) /n_{s,3}(0)=
n_s(T)/n_s(0),
\end{equation}
Here it will shown that for quisi 2D HTS (3) leads to
universal dependence of $T_c(\lambda^{-2}(0))$,
which was called "Uemura plot" and was discovered at
the measurements of muon relaxation rate [14].

For plane system (3) can be written as ratio
$ \rho_s(T/T_c)/ \rho_s(0)=n_s (T/T_c)/n_s (0)$, where
 $ \rho_s(T/T_c)$ is dimensionless hardness and satisfy
to  universal dependence [15]
\begin{equation}\label{4}
\rho_s(T/T_c)  = \exp (-\frac{T e^{-1}}{T_c\rho_s(T/T_c)})
\end{equation}
 The decision of (4) was received in ref.[15]:
$\rho_s(0)=1$, and at $T=T_c$
\begin{equation}\label{5}
\rho_s(T/T_c)|_{T=T_c} =e^{-1},
\end{equation}
 Expressions (3)-(4) results to universal
 character of temperature dependence of ratio
\begin{equation}\label{6}
\lambda^{2}(0) / \lambda^{2}(T/T_c)=
\rho_s(T/T_c)/ \rho_s(0)=
\exp(-\frac{T e^{-1}
\lambda^{2}(T/T_c)}{T_c \lambda^{2}(0)}),
\end{equation}
 and simple relation between values $\lambda^{2}(T)$
and $n_s (T)$ at  $T=T_c$ and at $T=0$:
\begin{equation}\label{7}
\lambda^{2}(0)/ \lambda^{2}(T_{c}) =
n_s (T_{á}) / n_s (0) = e^{-1}
\end{equation}
 Using this relation and Kosterlitz-Thouless-Nelson
formula [16]
\begin{equation}\label{9}
k_B T_c=\frac{ h^{2}}{ 32 \pi m} n_s(T_c),
\end{equation}
we can receive the universal relation between the density
 $n_s (0)$  at $T=0$ and transition temperature $T_c$:

\begin{equation}\label{10}
T_c=\frac{ h^{2} e^{-1}}{ 32 k_B \pi m} n_s(0),
\end{equation}
 where $k_B$ is Boltzman constant. For two-dimension
superconductor  the role of effective penetration
length acts magnetic screening length
\begin{equation}\label{11}
L_s (T)= \frac{mc^{2}}{2\pi n_s(T)\rm e^{2}},
\end{equation}
which is connected with the bulk London penetration
length
\begin{equation}\label{12}
     L_s = 2d^{-1} \lambda^{2},
\end{equation}
where $d$ - thickness of $Cu O_2$ plane.

 It is seen from (7-11) that the transition temperature is
proportional  to $\lambda ^{-2}(0)$
\begin{equation}\label{13}
T_c= k \lambda^{-2}(0),
\end{equation}
 where
\begin{equation}\label{14}
k= \frac{c^{2}  h^{2} e^{-1}d}{64 k_B \pi^{2} \rm e^{2}},
\end{equation}
depends only from universal constants and from thickness of
$Cu O_2$ plane. The temperature of dimensional crossover,
$T_{cr}<T_c$, depends on correlation length $ \xi _c (T)$,
and can be determined at the measurement of penetration
length as the boundary of two-dimensional region, where
at $T>T_{cr}$ the measurement values of
$\lambda^{2}(0)/\lambda^{2}(T/T_c)$ are
diverging from universal relation (6). Knowing $T_{cr}$
let us to determine the interval
$\Delta^{S}_{3D}\approx (T_c-T_{cr})$ of three-dimensional
fluctuations at $T<T_c$. Full interval is equal
\begin{equation}\label{15}
\Delta _{3D} = \Delta^{N}_{3D} + \Delta^{S}_{3D} = T^{0}_c -T_{cr}
\end{equation}
does not depend on the exactness of the measurement of $T_c$,
 and determines as the difference between the
temperatures of two  dimensional crossovers:
in normal state, $T^{0}_c$, and in superconducting
state, $T_{cr}$. So, for $La_{1.85} Sr_{0.15} Cu O_4$
the measurements of resistivity $\rho_c(T)$ [17] and
of the penetration length [18]
let us to determine
the temperature $T^{0}_c\sim 41.5 K$, and
$T_{cr} \sim 26 K$. This lead to the value
$\Delta _{3D}\sim 15.5 K$. We see that the
region $\Delta_{3D}$ of three-dimensional
superconducting fluctuations is finite in
the concrete, that evidences about the two-dimensional
character of superconducting transition.

Thus, here it is shown that expression (12), which was
found out at the measurements of muon relaxation rate [14]
for quisi 2D HTS, really is universal and is the sequence
of general consistent pattern (4), (8) of superconducting
state for two-dimensional systems and must to fulfil for
HTS with two-dimensional superconducting state at
$T<T_{cr}$. This means that for quisi 2D HTS superconducting
transition  has two-dimensional character and develops on
Kats scenario [7], and at $T<T_{cr}$ the temperature
dependence of $\lambda^{2}(0)/ \lambda^{2}(T/T_c)$ also
must to be universal (6).

The author would like to thank Dr.Kabanov V. for the reference on
the V.L.Pokrovskii paper [13].

{\bf References}

{\small1. S.L.Cooper, K.E.Gray, in "Physical properties of high
temperature  superconductors" ed.:

 Donald M.Ginsberg,IY p.61-188 (1994).

2. S.Martin, A.T.Fiory, R.M.Fleming, G.P.Espinoza, and
A.S.Cooper, Phys.Rev.Lett.

{\bf 62}, 677 (1989).

3.D.H.Kim, A.M.Goldman, J.H.Kang, R.T.Kampwirth, Phys.Rev.
{\bf B40}, 8834 (1989).

4. Y.Matsuda, S.Komiyama, T.Onogi, T.Terashima,
K.Shimura, and Y.Bando ,

Phys.Rev.{\bf B48}, 10498 (1993).

5. Y.Matsuda, S.Komiyama, T.Terashima, K.Shimura,
and Y.Bando, Phys.Rev.Lett.

{\bf 69}, 3228 (1992).

6. Z.Tesanovic, L.Xing, L.Bulaevskii, O.Li, and Suenaga,
Phys.Rev.Lett.{\bf 69}, 3563 (1992).

7. E.I.Kats, JE'F (Russian),{\bf 56}, 1675 (1969).

8. Vik.Dotsenko and M.V.Feigelman, JETP,{\bf 83}, 345 (1982).

9. L.Bulaevskii, Usp.Fiz.Nauk (Russian), {\bf 116}, 449 (1975).

10. G.Sergeeva, Physica C,{\bf 341-348}, 181 (2000).

11. G.G.Sergeeva, Low Temp. Phys.,{\bf 26}, 453 (2000);
cond-mat/0009212.

12. T.Schneider and H.Keler, Phys.Rev.{\bf B47}, 5915 (1993).

13. V.L.Pokrovskii, Pis'ma JETF, (Russian) {\bf 47}, 539 (1988).

14. Y.J.Uemura, V.J.Emery, A.R.Moodenbaugh
et al. Phys.Rev.{\bf B38}, 909 (1988);

Phys.Rev.Lett.{\bf 64}, 2082 (1990).

15. Patashinskii A.Z., Pokrovskii V.L.
Fluctuational theory of phase transition

(Russian, Moscow, Nauka) 1982.

16. B.I.Halperin and David R.Nelson, J. Low Temp.Phys.
{\bf 36}, 599 (1979).

17. T.Ito, H.Takagi, S.Ishibashi et al. Nature,
{\bf350}, 596 (1991).

18. G.Aeppli and R.J.Cava, E.J.Ansaldo et al.
Phys.Rev.{\bf B35}, 7129 (1987).}

\end{document}